\documentclass[nofootinbib,aip,graphicx]{revtex4-1}
\usepackage{graphicx}
\usepackage{epsfig}
\usepackage{amssymb}
\usepackage{amsmath}
\usepackage{epstopdf} 
\usepackage{subfloat}
\usepackage{lineno}
\usepackage{float}
\usepackage{color}
\usepackage{enumerate}
\usepackage{multirow}
\usepackage{rotating}
\usepackage{array}
\usepackage[version=4]{mhchem}
\usepackage{chemist}
\usepackage{tikz,chemfig}
\usepackage{blkarray}
\usepackage{gensymb}
\usepackage{amsmath,amssymb,mathrsfs,bm,amsthm}
\usepackage{array}
\usepackage{mathtools}
\usepackage{float}
\usepackage{ulem}
\usepackage{siunitx}
\usepackage{bbold}
\usepackage[dvipsnames]{xcolor}

\usepackage{fixltx2e}

\usepackage{pgfplots}
\pgfplotsset{compat = newest}

\usepackage{comment}

\usepackage{pgfplots}
\pgfplotsset{compat = newest}
\usepackage{tikz}
\usetikzlibrary{calc}
\usepackage{subcaption}
\usetikzlibrary{arrows}
\usetikzlibrary{arrows.meta}
\usetikzlibrary{positioning}			
\usetikzlibrary{calc}				
\usetikzlibrary{matrix}

\newcommand{\cAA}[1]{\textcolor{black}{#1}}

\pgfplotsset{
  log x ticks with fixed point/.style={
      xticklabel={
        \pgfkeys{/pgf/fpu=true}
        \pgfmathparse{exp(\tick)}%
        \pgfmathprintnumber[fixed relative, precision=3]{\pgfmathresult}
        \pgfkeys{/pgf/fpu=false}
      }
  },
  log y ticks with fixed point/.style={
      yticklabel={
        \pgfkeys{/pgf/fpu=true}
        \pgfmathparse{exp(\tick)}%
        \pgfmathprintnumber[fixed relative, precision=3]{\pgfmathresult}
        \pgfkeys{/pgf/fpu=false}
      }
  }
}

\begin{document}

\title{Systematically Improved Potential Energy Surfaces via sinNN Models and Sparse Grid Sampling}
\author{Antoine Aerts} 
\email{antoine.aerts@ulb.be}
\affiliation{Universit\'e libre de Bruxelles, Spectroscopy, Quantum Chemistry and Atmospheric Remote Sensing (SQUARES), 50, av. F. Roosevelt CP 160/09, 1050 Brussels, Belgium}

\begin{abstract}
    {\large{\textbf{Abstract}}}
    \linebreak
    Accurate, global Potential Energy Surfaces (PES) expressed in sum-of-products (SOP) form are a prerequisite for efficient high-dimensional quantum dynamics simulations using the Multi-Configuration Time-Dependent Hartree (MCTDH) method. 
This work introduces a methodology for constructing such surfaces by combining hierarchical sparse grid sampling with a single-layer neural network using sinusoidal activation functions (sinNN). 
The sparse grid strategy provides a rigorous, unbiased discretization of the configuration space, enabling systematic improvability of the PES fidelity, where accuracy is strictly controlled by the refinement level, while successfully mitigating the curse of dimensionality. 

The sinNN fitting approach leverages a trigonometric factorization identity to maintain a compact SOP form, offering superior numerical stability compared to ``standard'' exponential-based networks (expNN) for the molecular systems investigated. 
We validate this framework by refitting an analytical PES for nitrous acid (HONO). 
The flexibility of the sparse grid methodology is demonstrated through a dual-reference strategy, where grids centered on distinct isomers are merged to eliminate topological bias. 
This optimized sampling yields a global PES that reproduces fundamental vibrational transition energies for both \textit{trans}- and \textit{cis}-HONO with spectroscopic precision ($<$ 2.5~cm$^{-1}$) and high data efficiency. 

Finally, the methodology is applied to fit potential energies computed via the AI-enhanced quantum mechanical method AIQM2. 
The resulting AIQM2-based PES for HONO reproduces experimental vibrational frequencies with a root mean square deviation of $\sim$16~cm$^{-1}$, a performance comparable to high-level \textit{ab initio} methods. 
The robustness of the approach is further confirmed on larger molecules, formic acid (HCOOH) and carbamic acid (H$_2$NCOOH), establishing the combination of sparse grid sampling and sinNN fitting as a powerful, automated tool for generating topologically sound, spectroscopic-quality potential energy surfaces.

\end{abstract}
\maketitle

\maketitle

This article may be downloaded for personal use only. Any other use requires prior permission of the author and AIP Publishing. This article appeared in Antoine Aerts; Systematically improved potential energy surfaces via sinNN models and sparse grid sampling. J. Chem. Phys. 28 March 2026; 164 (12): 124306; and may be found at \url{https://doi.org/10.1063/5.0320172}.
\section{Introduction}

Vibrational structure and quantum dynamics simulations are indispensable for elucidating molecular behavior in fields ranging from high-resolution spectroscopy to reaction dynamics.\cite{wircms_12_e1605,book_bowman} Central to these simulations is the Born-Oppenheimer potential energy surface (PES),\cite{bor27:457} which maps the electronic energy as a function of nuclear geometry. While the electronic Schr\"odinger equation provides potential energies at static geometries, the subsequent solution of the time-dependent nuclear Schr\"odinger equation ideally requires a global, analytical representation of the PES landscape.

A fundamental requirement for any modern PES construction methodology is the ability to preserve the intrinsic accuracy of the underlying electronic structure calculations used to compute the potential energies. 
The fitting procedure should not act as a bottleneck that degrades high-level \textit{ab initio} data; rather, it should be systematically improvable, allowing the user to tailor the convergence of the PES to the specific demands of the application. 

To achieve this tunable fidelity without rendering the problem computationally intractable, a PES methodology must satisfy a stringent set of criteria: 
(1) Systematic Improvability: Both sampling density and fitting accuracy must be hierarchically scalable to converge towards the true potential energy landscape.
(2) Sampling Efficiency: Training points must be information-rich to minimize the cost of high-level quantum chemistry.
(3) Compact SOP Form: To enable efficient high-dimensional quantum propagation, particularly within the Multi-Configuration Time-Dependent Hartree (MCTDH) framework,\cite{man92:3199,bec00:1,mey03:251} the PES must be strictly maintained in a sum-of-products (SOP) form.
(4) Universality: The model must be flexible enough to capture diverse topological features and accurate enough to reproduce training data without unphysical artifacts.
(5) Accessibility: The approach should rely on existing tools and be easy to implement.
(6) Data Efficiency: The method must be robust enough to train from sparse datasets.

Existing approaches often force a compromise between these goals. Direct dynamics (``on-the-fly'') methods avoid pre-fitting but are computationally prohibitive, often restricting the user to lower-level electronic structure theory.\cite{pccp_26_1829} 
\cAA{Within the vast literature on PES fitting, machine learning methods have emerged as powerful tools capable of providing accurate specialized, and general models for both small and large systems. \cite{man20:10187,beh21:10037,mlatom3}
However, leveraging the efficiency of the MCTDH method imposes the PES to be written in SOP form,\cite{wan15:7951,mctdh:package,mey90:73} where each term is a product of single-degree-of-freedom factors. 
Manzhos and Carrington\cite{man06:194105} addressed this by demonstrating that a SOP PES can be obtained from neural networks (NNs) via the ``expNN'' method based on exponential activation functions. This approach has since been extended to focus on the compactness of high-dimensional models\cite{man07:014103,man08:224104} and further generalized by Koch and Zhang,\cite{koc14:021101} who introduced multiplicative neurons to allow for diverse activation functions while retaining the SOP structure.}

\cAA{Beyond neural network representations, significant progress in automated SOP construction has been achieved through tensor decomposition techniques. Peláez and coworkers have introduced a series of increasingly efficient methods for the HONO system, moving from the multigrid POTFIT (MGPF) approach\cite{pel13:014108} to analytical representations in the Tucker (SOP-FBR)\cite{pan20:234110} and Canonical Polyadic (CP-FBR)\cite{nad23:114109} forms. These methods prioritize the compactness of the Hamiltonian operator, often requiring low-rank optimizations to achieve high accuracy.}

\cAA{In this work, we explore the utility of combining hierarchical sparse grid sampling with single-layer NN fitting to create a methodology that aims to be systematically improvable and capable of reflecting the accuracy of the chosen quantum chemistry method. 
We propose a ``sinNN'' approach that replaces exponential activation functions with sinusoidal ones, retaining the crucial SOP structure while leveraging the universal approximation capabilities of NNs. 
By coupling this with sparse grid sampling,\cite{bungartz2004sparse} we hope to facilitate a hierarchical and unbiased exploration of the PES. 
However, we must acknowledge that NNs are notoriously data-hungry and highly sensitive to the specific distribution of their training sets. 
Furthermore, while the nested nature of sparse grids provides a clear path for systematic improvement, the specific grid type chosen in this work is unlikely to be optimal for the topographies of molecular PESs, especially as we are assessing the performance of our PES models by computing fundamental bound vibrational states energies. 
For these reasons, we anticipate a poor economy of points, where a high density of samples is required to achieve the desired accuracy.}

Anharmonic vibrational transition energies are computed using the improved relaxation method implemented in the Heidelberg MCTDH package,\cite{mey06:179,dor08:224109} a rigorous validation metric that relies directly on the SOP form of the fitted Hamiltonian. 
We apply this methodology to three semi-rigid, covalently bound systems of increasing complexity: nitrous acid (HONO), formic acid (HCOOH), and carbamic acid (H$_2$NCOOH). 
HONO serves as a benchmark for \textit{trans} to \textit{cis} isomerization,\cite{franco2024pyrogenic,ric04:1306} while HCOOH and H$_2$NCOOH specifically introduce challenges associated with their larger skeletal frames.

Additionally, we illustrate the methodology's versatility through two distinct scenarios: the refitting of an established analytical PES and the direct fitting of energies obtained from the machine learning-enhanced semi-empirical method AIQM2.\cite{mlatom3,Chen2024-qa} 
This dual approach allows to evaluate the intrinsic fitting efficiency of the sinNN models as well as the capability of AIQM2 to describe PES landscapes beyond equilibrium geometries.

The remainder of this paper is organized as follows. Section~\ref{sec:methods} details the sparse grid sampling and sinNN fitting protocols, with a specific focus on the model parameter initialization strategy.
Section~\ref{sec:benchmark_refit} benchmarks the intrinsic accuracy of the sinNN models against established expNN approaches by refitting the analytic Richter PES for HONO; this section includes a critical evaluation of sampling bias and introduces an optimized dual-reference sampling strategy. 
Section~\ref{sec:aiqm2_pes} applies the validated methodology to potential energies computed via the AIQM2 method.
This section validates the quality of the generated PES against experimental data for HONO and assesses the topological robustness of the method when applied to the larger formic acid and carbamic acid systems. 
Finally, Section~\ref{sec:concl} provides concluding remarks and perspectives on future improvements.

\section{Methods}
\label{sec:methods}
\subsection{Sum-of-products form using single-layer artificial neural networks: from expNN to sinNN fitting}

The fitting methodology presented here, termed ``sinNN'', is a natural extension of the ``expNN'' approach introduced by Manzhos and Carrington.\cite{man06:194105} It retains the core advantage of expNN: the ability to generate a closed-form SOP PES directly from scattered data, without requiring an intermediate grid or auxiliary interpolation steps.

\subsubsection{expNN}

The expNN method utilizes a feed-forward neural network with a single hidden layer of exponential activation functions. 
Given a molecular geometry $\mathbf{x} = (x_1, \dots, x_D)$ and its corresponding energy $V(\mathbf{x})$, the network output is defined as:

\begin{equation}
    V^\text{expNN}(\mathbf{x}) = \sum_{i=1}^{N} w_i^{(2)} \exp\left(\sum_{j=1}^{D} w_{ij}^{(1)} x_j + b_i^{(1)}\right)+b_i^{(2)},
\end{equation}

where $w_{ij}^{(1)}$ are the input-layer weights mapping coordinate $x_j$ to the $i$-th neuron, $b_i$ are biases, and $w_i^{(2)}$ are weights connecting the hidden layer to the output node. 
By algebraically expanding the exponential of the sum into a product of exponentials, the biases $b_i^{(2)}$ and scaling factors can be absorbed into effective coefficients $c_i$, yielding the required SOP form:

\begin{equation}
    V^\text{expNN}(\mathbf{x}) = \sum_{i=1}^{N} c_i \prod_{j=1}^{D} \exp(w_{ij}^{(1)} x_j) + V_0.
\end{equation}
The model has a total of $N(D+2)+1$ adjustable parameters.

\subsubsection{sinNN}

The sinNN method replaces the unbounded exponential activation with a periodic sinusoidal function:
\begin{equation}
V^{\text{sinNN}}(\mathbf{x}) = \sum_{i=1}^{N} w_i^{(2)} \sin\left( \sum_{j=1}^{D} w_{ij}^{(1)} x_j + b_i^{(1)} \right) + b^{(2)}. \label{eq:sinNN}
\end{equation}
While this preserves the parameter count of expNN ($N(D+2)+1$), the transformation to SOP form is non-trivial. Standard trigonometric expansion of $\sin(\sum \theta_j)$ generates $2^{D-1}$ terms per neuron, which scales exponentially with dimensionality $D$. To avoid this combinatorial explosion, we employ the identity derived by Mohlenkamp and Monzón:\cite{moh05:65}
\begin{equation}
\sin\left( \sum_{j=1}^{D} x_j \right) = \sum_{j=1}^{D} \sin(x_j) \prod_{\substack{k=1 \ k \neq j}}^{D} \frac{\sin(x_k + \alpha_k - \alpha_j)}{\sin(\alpha_k - \alpha_j)}.
\end{equation}
This identity reduces the separation rank from $2^{D-1}$ to exactly $D$. Consequently, the sinNN potential can be expressed as a compact SOP:
\begin{equation}
V^{\text{sinNN}}(\mathbf{x}) = \sum_{i=1}^{N} w_i^{(2)} \sum_{j=1}^{D} \sin\bigl( w_{ij}^{(1)} x_j + b_i^{(1)} \bigr) \prod_{\substack{k=1 \ k \neq j}}^{D} \frac{\sin\bigl( w_{ik}^{(1)} x_k + b_i^{(1)} + \alpha_k - \alpha_j \bigr)}{\sin(\alpha_k - \alpha_j)} + b^{(2)},
\end{equation}
where $\alpha_j$ are free structural parameters. This formulation scales linearly with $D$ (complexity $O(ND)$), making sinNN computationally tractable for high-dimensional systems.
Beyond computational efficiency, the sinusoidal form offers a physical advantage. The weights and biases in Eq.~\ref{eq:sinNN} essentially map to the frequencies, phases, and amplitudes of a Fourier-like expansion. Since molecular potentials (especially multi-well landscapes) are bounded and often possess periodic or quasi-periodic symmetries, a periodic basis is arguably a more natural approximator than the monotonic exponentials of expNN.\cite{cybenko1989approximation}

\subsection{Network parameters initialization and training}
\label{sec:trainingmethod}
The non-convex optimization of neural networks is highly sensitive to parameter initialization.\cite{narkhede2022review} To mitigate saturation effects and ensure efficient training, we implement a modified Nguyen-Widrow initialization scheme.\cite{nguyen1990improving}
The input weights $w_{ij}^{(1)}$ and output weights $w_{i}^{(2)}$ are initially drawn from a uniform distribution $\mathcal{U}(-0.5, 0.5)$. To ensure that neurons are evenly distributed across the input domain, the input weight vectors are first normalized and then scaled by a factor $\beta$:
\begin{equation}
\mathbf{w}_i^{(1)} \leftarrow \beta \frac{\mathbf{w}_i^{(1)}}{\lVert\mathbf{w}_i^{(1)}\rVert}, \quad \text{with} \quad \beta = 0.7 N^{\frac{1}{D}}.
\end{equation}
Unlike the standard Nguyen-Widrow method, which distributes biases uniformly across $[-\beta, \beta]$, we initialize biases $b_i^{(1)}$ with small random values $\mathcal{U}(-0.1, 0.1)$. This prevents symmetry-breaking issues during the early stages of optimization. Finally, output weights are normalized to $\lVert \mathbf{w}^{(2)}\rVert = 1$ to maintain a uniform output magnitude at the start of training.
Training is performed using the Levenberg-Marquardt (LM) algorithm,\cite{mar63:431} a second-order optimization method that interpolates between the Gauss-Newton algorithm and gradient descent. The LM update rule is given by:
\begin{equation}
(J^T J + \mu I) \Delta \boldsymbol{w} = -J^T \mathbf{r},
\label{eq:jacobian}
\end{equation}
where $J$ is the Jacobian matrix of the residuals $\mathbf{r}$, and $\mu$ is a damping parameter. The LM algorithm is particularly well-suited for this application as it offers rapid convergence for regression problems with moderate parameter counts, avoiding the slow plateaus typical of first-order gradient descent.
Data is pre-processed by normalizing inputs and outputs to the $[0, 1]$ interval. To prevent overfitting, the dataset is partitioned into training, validation, and test sets. We employ an early stopping criterion based on the test set Root Mean Square Error (RMSE):
\begin{equation}
\text{RMSE} = \sqrt{\frac{1}{m}\sum_{p=1}^m \left(V^\text{model}(\mathbf{x}_p)-V^\text{ref}(\mathbf{x}_p)\right)^2}.
\end{equation}
\cAA{Specifically, training is terminated if the test set error fails to decrease for 50 consecutive iterations. 
This ensures that the optimization concludes before the model begins to overfit the training data. 
The validation set remains entirely sequestered throughout this process; it is never ``seen'' by the training algorithm or used to determine the stopping point. 
It is evaluated only once after the training has fully converged, aiming to provide a strictly independent benchmark of the model’s performance.}

\subsection{Molecular geometries sampling with sparse grids}
\label{sec:SG}
Sparse grids\cite{bungartz2004sparse} provide an efficient means of representing and computing high-dimensional functions while mitigating the curse of dimensionality. Unlike full tensor-product grids, which scale exponentially with dimensionality, sparse grids leverage hierarchical refinements to balance accuracy and computational efficiency of sampling tasks. The position of points in a sparse grid depend on indices, and it is illustrated in Figure~\ref{fig:SG2D} up to two dimensions. 

\begin{figure}[h!]
\captionsetup[subfigure]{labelformat=empty}
		\centering
  		\begin{subfigure}{0.15\textwidth}
		\centering
	\begin{tikzpicture}
    \begin{axis}[
        width=3.75cm,
        height=3.75cm,
        xmin=0,xmax=1,
    ymin=0,ymax=1,
    xtick={\empty},ytick={\empty},]
    \addplot[matrix plot, point meta=explicit]
        coordinates {}; 
        \node at (0.5,0.5)  {$+$};
    \end{axis}
\end{tikzpicture}
		\caption{(1,1)}
		\end{subfigure}%
		\hfill
		\begin{subfigure}{0.15\textwidth}
		\centering
\begin{tikzpicture}
    \begin{axis}[
        width=3.75cm,
        height=3.75cm,
        xmin=0,xmax=1,
    ymin=0,ymax=1,
    xtick={\empty},ytick={\empty},]
    \addplot[matrix plot, point meta=explicit]
        coordinates {
            (0.25,0.5) [0.] (0.75,0.5) [1.] 
            
            (0.25,1.5) [NaN] (0.75,1.5) [Nan] 
            
        }; 
        \node at (0.25,0.5)  {$+$};
        \node at (0.75,0.5)  {$+$};
    \end{axis}
\end{tikzpicture}
		\caption{(2,1)}
		\end{subfigure}%
		\hfill
		\begin{subfigure}{0.15\textwidth}
		\centering
	\begin{tikzpicture}
    \begin{axis}[
        width=3.75cm,
        height=3.75cm,
        xmin=0,xmax=1,
    ymin=0,ymax=1,
    xtick={\empty},ytick={\empty},]
    \addplot[matrix plot, point meta=explicit]
        coordinates {
            (0.5,0.25) [0.] (0.5,0.75) [1.] 
            
            (1.5,0.25) [NaN] (1.5,0.75) [Nan] 
            
        }; 
        \node at (0.5,0.25)  {$+$};
        \node at (0.5,0.75)  {$+$};
    \end{axis}
\end{tikzpicture}
		\caption{(1,2)}
		\end{subfigure}%
		\hfill
		\begin{subfigure}{0.15\textwidth}
		\centering
	\begin{tikzpicture}
    \begin{axis}[
          width=3.75cm,
        height=3.75cm,
        xmin=0,xmax=1,
    ymin=0,ymax=1,
    xtick={\empty},ytick={\empty},]
    \addplot[matrix plot, point meta=explicit]
        coordinates {
            (0.25,0.75) [0.] (0.75,0.75) [1.] 
            
            (0.25,0.25) [0.25] (0.75,0.25) [0.5] 
            
        }; 
        \node at (0.25,0.25)  {$+$};
        \node at (0.75,0.75)  {$+$};
        \node at (0.25,0.75)  {$+$};
        \node at (0.75,0.25)  {$+$};
    \end{axis}
\end{tikzpicture}
		\caption{(2,2)}
		\end{subfigure}%
		\hfill
  		\begin{subfigure}{0.15\textwidth}
		\centering
  \begin{tikzpicture}
    \begin{axis}[
        width=3.75cm,
        height=3.75cm,
        xmin=0,xmax=1,
    ymin=0,ymax=1,
    xtick={\empty},ytick={\empty},]
    \addplot[matrix plot, point meta=explicit]
        coordinates {
            (0.125,0.5) [0.] (0.375,0.5) [0.25] (0.625,0.5) [0.5] (0.875,0.5) [1.]
            
            (0.125,1.5) [NaN] (0.375,1.5) [NaN] (0.625,1.5) [NaN] (0.875,1.5) [NaN]
            
        }; 
        \node at (0.125,0.5)  {$+$};
        \node at (0.375,0.5)  {$+$};
        \node at (0.625,0.5)  {$+$};
        \node at (0.875,0.5)  {$+$};
    \end{axis}
\end{tikzpicture}
		\caption{(3,1)}
		\end{subfigure}%
		\hfill
    		\begin{subfigure}{0.15\textwidth}
		\centering
  \begin{tikzpicture}
    \begin{axis}[
        width=3.75cm,
        height=3.75cm,
        xmin=0,xmax=1,
    ymin=0,ymax=1,
    xtick={\empty},ytick={\empty},]
    \addplot[matrix plot, point meta=explicit]
        coordinates {
            (0.125,0.75) [0.] (0.375,0.75) [1.]  (0.625,0.75) [0.125] (0.875,0.75) [0.375]
            
            (0.125,0.25) [0.25] (0.375,0.25) [0.5] (0.625,0.25) [0.625] (0.875,0.25) [0.875]
        }; 
        \node at (0.125,0.75)  {$+$};
        \node at (0.375,0.75)  {$+$};
        \node at (0.625,0.75)  {$+$};
        \node at (0.875,0.75)  {$+$};
        \node at (0.125,0.25)  {$+$};
        \node at (0.375,0.25)  {$+$};
        \node at (0.625,0.25)  {$+$};
        \node at (0.875,0.25)  {$+$};
    \end{axis}
\end{tikzpicture}
		\caption{(3,2)}
		\end{subfigure}%

		\caption{Position of points on (regular) sparse grids within the unit square given grid indices $(l_1, l_2)$. The reference index is 1, increasing this index further subdivides the space spanned by the points. Colors are arbitrary, but emphasize the local span of each point, that gets smaller with increased index level.}
		\label{fig:SG2D}
\end{figure}

Consider a function $f(x)$ defined on the unit interval $[0, 1]$. A standard full grid of level $t$ comprises $2^t + 1$ equidistant points, sampling the domain uniformly. In contrast, the sparse grid used here is built from a complete, hierarchical set of \cAA{(equidistant)} nested points. By exploiting this hierarchical structure, the sparse grid can represent multidimensional functional variations with an accuracy comparable to that of a full grid, but using significantly fewer points overall. At level $l>1$, the grid consists of $2^{l-1} + 1$ points but includes only those necessary to refine coarser levels, forming a hierarchical adaptive structure. Taking a regular grid where edge positions are excluded, new positions within level $l$, \textit{i.e.}, excluding points of lower level grids, are obtained with:

\begin{equation}
X_l = \left\{ x_i = \frac{i}{2^l} \mid i = 1, \dots, 2^l, \;i\;\text{odd}\right\}.
\end{equation}

Extending sparse grids to more dimensions within $[0,\;1]^D$, a standard full tensor-product grid of level $t$ requires $(2^t + 1)^D$ points. Sparse grids, in contrast, are constructed by summing lower-dimensional tensor products:

\begin{equation}
\text{SG}_t = \bigcup_{l_1 + \ldots +l_D \leq t} X_{l_1}\times\ldots \times X_{l_D},
\end{equation}

where $l_1$, $\ldots$, $l_D$ are the refinement levels in each dimension. This construction has been shown to significantly reduce the number of grid points required while preserving accuracy to capture function variations, particularly for smooth ones.\cite{bungartz2004sparse}

\cAA{Avila and Carrington~\cite{avi15:044106} were first to introduce sparse grids for PES model construction, combining sampling with an interpolation scheme. 
Because they refit a known PES model with an interpolative approach, they could properly limit sampling to grid levels that directly contributed to the model's reproduction. 
In contrast, the sinNN approach is a fitting method, making such constraints tentative and subject to validation; however, we can still restrict the maximum grid rank as will be discussed later.}

\cAA{Several alternative strategies exist for constructing informative training sets. 
Previous expNN fits (see Supplementary Material for an overview) have utilized fitting sets based on: grids,\cite{pra13:69225} random sampling,\cite{man06:194105,man08:224104} Sobol sequences,\cite{MANZHOS2023355} hybrid approaches combining grids and random points,\cite{pra16:158,pra16:174305,pccp_19_22272,bro17:1730001,pra20:e1674936} and trajectories.\cite{man08:224104} 
These selections are frequently guided by predictive filters designed to favor low-energy configurations and avoid sampling high-energy regions. 
Common approaches include trajectory sampling,\cite{bra09:577,qu18:151} and adaptive or active learning schemes that iteratively select points based on model uncertainty,\cite{dra17:244108,hua25:035004,sch23:144118} error estimation,\cite{aerts2022} or explicit wavefunction density criteria.\cite{jcp_136_224105,ric14:1401,sch20:064105,art20:194105}}

\cAA{Recently, Schneider et al.\cite{sch23:144118} systematically compared static and Gaussian process regression-based adaptive schemes for positioning \textit{ab initio} points to construct $n$-mode PESs. 
They concluded that while adaptive sampling reduces the number of required points, the exact positioning of grid points is of secondary importance; simpler static or gradient-weighted schemes performed as well as more sophisticated approaches for many systems. 
In contrast, our work prioritizes the homogeneous, hierarchical structure of sparse grids, which guarantees systematic coverage and provides a clear, unbiased path to convergence, a core criteria of our methodology.}
  
\cAA{Avila and Carrington employed nested Chebyshev points, but other hierarchical configurations have been explored, such as Smolyak grids using Leja point selection~\cite{avi17:064103,wod19:154108} or nested 1-D sequences~\cite{wod21:114107} to optimize interpolation accuracy. 
While we prioritize homogeneous sampling, we acknowledge that alternative grid types, such as Chebyshev or Leja points, should be superior for vibrational transition energies. 
Consequently, our choice of grid while providing a systematic path for improvement, is not optimal for the PES topography, leading to an anticipated poor economy of points.}

\subsection{Energy Filter from Low-Dimensional Interpolation}
\label{sec:energy_filter}

To \cAA{mitigate computational effort}, we implement a predictive energy filter that prunes high-energy configurations before the expensive \textit{ab initio} calculation. 
A low-cost interpolation model is constructed using only grid points from low \cAA{hierarchical levels ($l \leq 4$) and with a coupling dimensionality $n \leq 2$, meaning that the basis functions correlate at most two coordinates at a time.} 
Points exceeding a predefined energy threshold are excluded. 
The interpolation is defined as a linear combination of tensor-product basis functions, where each function is centered at a specific grid point and scaled according to its hierarchical level. 
The interpolation is expressed as
\begin{equation}
V(x,y) = \sum_{(l_x,l_y), (j_x,j_y)} d_{(l_x,l_y), (j_x,j_y)} \prod_{i=x,y} \phi_{l_i,j_i}(x_i),
\end{equation}
where $d_{(l_x,l_y), (j_x,j_y)}$ represents the model coefficients, $l_i$ denotes the grid's hierarchical level, $j_i$ represents a point index, and $\phi_{l_i,j_i}(x_i)$ are one-dimensional basis functions with local support, defined as
\begin{equation}
\phi_{l,j}(x) = \begin{cases} \displaystyle \prod_{k=0}^p \frac{x - x_k}{x_{i,j}-x_k} & \text{if } |x - x_{i,j}| < h_i \\ 0 & \text{otherwise}\end{cases},
\end{equation}
where $x_k$ are the positions of the grid point ``ancestors'', \textit{i.e.}, the points that $x_{i,j}$ refines, and half its support width is equal to $h_i$. 
\cAA{In this scheme, increasing the level $l$ of the hierarchical basis functions provides a refinement over the lower-level ones, and the $d$ coefficients are therefore known as the hierarchical surplus. They are calculated as the difference between the sampled function value at the new point $x_{i,j}$ and the value predicted by the basis functions of its ancestor points $x_k$.}
By limiting the interpolation to points that belong to grid levels up to $l=4$, the method maintains a balance between computational efficiency and representation accuracy of the PES appropriate to its filtering purpose.

\subsection{Kinetic Energy Operators from TANA}
\begin{figure}
    \centering
    \subfloat[]{\includegraphics[width=0.45\columnwidth]{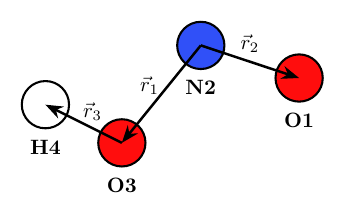}}
    \subfloat[]{\includegraphics[width=0.35\columnwidth]{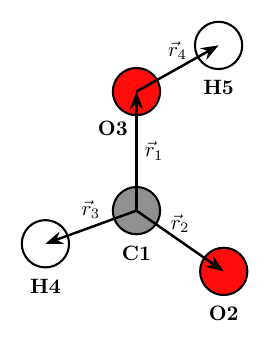}}\\
    \subfloat[]{\includegraphics[width=0.5\columnwidth]{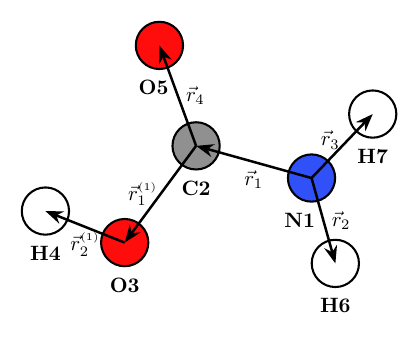}}
    \caption{Molecular structures and corresponding valence vector coordinate systems for (a) HONO, (b) HCOOH, and (c) H$_2$NCOOH. Vectors $\vec{r}_j^{\scriptscriptstyle(i)}$ belong to the coordinates sub-system $(i)$.}
    \label{fig:valence_coordinates}
\end{figure}

Exact analytical KEOs were derived using the TANA program.\cite{ndo12:034107,ndo13:204107} TANA automates the polyspherical approach,\cite{gat09:1} generating the KEO directly in the SOP form required by MCTDH. 
The internal coordinates (illustrations for the chosen systems in Figure~\ref{fig:valence_coordinates}) are defined via user-specified internal vectors, \textit{e.g.}, Jacobi, Radau, or valence vectors. 
TANA also handles the exact transformation between Cartesian and internal curvilinear coordinates, facilitating the direct mapping of the sparse grid samples to the quantum chemistry input.

Deriving an analytical KEO in SOP form for high-dimensional systems using curvilinear coordinates has long been a significant challenge, often requiring laborious and error-prone manual algebra. The formalization of the polyspherical approach provided a systematic theoretical framework for this task.\cite{gat09:1} The development of the TANA program, which automates this framework,\cite{ndo12:034107,ndo13:204107} represents a crucial methodological advance.

\subsection{The AI-enhanced Quantum Mechanical Method MLatom/AIQM2}
MLatom/AIQM2\cite{mlatom3,Chen2024-qa} is a general-purpose AI-enhanced quantum mechanical approach that aims to achieve coupled cluster (CCSD(T)/CBS) accuracy at the cost of semi-empirical methods. It employs a delta-learning framework to correct the GFN2-xTB baseline\cite{jctc_15_1652} with machine learning predictions from an ensemble of ANI neural networks,\cite{smith2017ani,gao2020torchani} further refined through transfer learning. 

While AIQM2 has demonstrated high accuracy for reaction barriers and thermochemistry,\cite{natc_12_7022} its performance for global PES construction remains less explored, particularly in regions away from equilibrium. This work specifically evaluates the ability of AIQM2 to support full-dimensional vibrational spectroscopy, testing its robustness when applied to the anharmonic regions of the PES.

\section{Results and Discussion}


The evaluation of the proposed methodology is conducted in two distinct phases. 
First, in Section~\ref{sec:benchmark_refit}, we assess the intrinsic numerical accuracy of the fitting procedure by refitting an established analytical PES for HONO.\cite{ric04:1306} 
This benchmark isolates the numerical fitting error from the intrinsic inaccuracies of the electronic structure method.

Second, in Section~\ref{sec:aiqm2_pes}, we apply the methodology to construct new PES models using potential energies computed via the AIQM2 method. 
The spectroscopic accuracy of the resulting HONO surface is examined against experimental data. 
Subsequently, we assess the robustness of the approach when applied to larger systems: formic acid (\ce{HCOOH}) and carbamic acid (\ce{H2NCOOH}). 
For these larger systems, the analysis focuses on the ground vibrational state to verify that the generated landscapes are topologically sound and free of unphysical artifacts.

\subsection{Benchmarking the sinNN Methodology: Refitting Richter's HONO PES}
\label{sec:benchmark_refit}

We first consider the refitting of a PES model for the small covalently bound HONO, with emphasis on the reproduction of computed infrared signatures. 
The PES model of Richter and coworkers\cite{ric04:1306} is considered a high-accuracy benchmark for the ground electronic state of HONO, covering both \textit{trans} and \textit{cis} isomers. 
Richter \textit{et al.} fitted a sum-of-products polynomial expansion to CCSD(T)/cc-pVQZ(-g) \textit{ab initio} potential energies and validated the PES quality by comparing vibrational transition energies with experimental values.

The internal curvilinear valence coordinates used in this work are illustrated in Figure~\ref{fig:valence_coordinates}. 
Vibrational motion is described explicitly in terms of these nuclear coordinates: stretching modes ($R_\text{OH}$, $R_\text{NO}$, $R_\text{N=O}$), bending angles ($\theta_\text{ONO}$, $\theta_\text{HON}$), and the out-of-plane torsion ($\tau$). 

\cAA{Table~S2 of the Supplementary Material} lists the fundamental vibrational transition energies of \textit{trans}- and \textit{cis}-HONO obtained from the Richter PES using the improved relaxation method. 
These values serve as the reference ``truth'' for our fitting benchmark. 
Computed energies are not strictly identical to those reported by Richter and coworkers in their original work due to the different versions of the MCTDH code and improved relaxation parameters used. 
The parameters were chosen such that the vibrational eigenstates energies are converged below the reported digits. 
Parameters are kept identical throughout this work and provided in the Supplementary Material.

\subsubsection{Sampling Strategy}
\label{sec:gentraining}

The sampling procedure constructs sparse grids within the internal curvilinear valence coordinates of HONO. 
These coordinates were selected for three primary reasons: they correspond directly to those used in the reference Richter PES; they match the KEO used in dynamics simulations; and they minimize high-order mode--mode coupling terms, facilitating compact PES representations.\cite{jcp_136_224105,pccp_26_11469}

The grid boundaries match the validity range of the Richter model. 
To maximize data efficiency, a critical optimization was the truncation of sampling dimensionality. 
While a full-dimensional treatment suggests 6D grids, sampling was restricted to a maximum rank of 4. 
This is justified because the reference Richter PES is a 4-body expansion; higher-dimensional sampling would be redundant. 
Furthermore, for semi-rigid molecules, 5- and 6-body interactions typically make negligible energetic contributions.\cite{yu22:296}

A dataset was generated using the hierarchical sparse grid approach centered on the \textit{trans}-HONO equilibrium geometry. 
We applied \cAA{the predictive energy filter, as defined in Section~\ref{sec:energy_filter}}, to prune points exceeding 10,000~cm$^{-1}$ relative to the global minimum. 
This strategy effectively focused computational resources on the physically relevant regions, resulting in a final master dataset of 266,478 points. 
The potential energy distribution (Figure~\ref{fig:dataset300k_full}) indicates a thorough exploration of the landscape, though strictly homogeneous sampling yields a distribution peak around 8,000~cm$^{-1}$. 
The resulting dataset maintains a well-balanced hierarchical structure, where the number of points scales with the dimensionality of the coupling terms: the count of 1D points is significantly lower than that of 2D, which is smaller than 3D, with the largest population found in 4D.

\begin{figure} 
\centering 
\resizebox{\textwidth}{!}{
        \input{4D_300k_15000.pgf}} 
\caption{Distribution of sampled potential energies relative to the global minimum of the \textit{trans}-HONO isomer. The data is categorized by the dimensionality of the coordinate variations (1D to 4D) relative to the reference equilibrium geometry. The dataset was generated via hierarchical sparse grid sampling on the Richter PES.} 
\label{fig:dataset300k_full} 
\end{figure}

\subsubsection{Convergence Behavior and Model Compactness}
\label{sec:Reprod_models}

To rigorously evaluate systematic improvability, we constructed a series of sinNN and expNN models with up to $N=70$ neurons. 
Models were trained on hierarchical subsets of the master dataset (25k, 50k, and 75k points), representing successive levels of grid refinement. 
Each subset was partitioned into training (80\%), test (10\%), and validation (10\%) sets. 
An energy filter was subsequently applied to remove points above 10,000~cm$^{-1}$. 
As shown in Table~\ref{tab:dataset_breakdown}, this protocol yields statistically consistent datasets that preserve the 80/10/10 ratio, ensuring that the validation metrics remain comparable across hierarchy levels.

\begin{table}[H]
    \centering
    \caption{Composition of the hierarchical datasets used to refit the HONO PES after applying a 10,000 cm$^{-1}$ energy filter. The subsets are labeled according to the number of points in the original sparse grid.}
    \label{tab:dataset_breakdown}
    \begin{tabular}{l c c c c}
        \hline
        \hline
        \textbf{Dataset} & \textbf{Total Points} & \textbf{Training} & \textbf{Test} & \textbf{Validation}  \\
        \hline
        25k & 23,238 & 18,590 & 2,348 & 2,329  \\
        50k & 45,671 & 36,534 & 4,692 & 4,629  \\
        75k & 67,687 & 54,169 & 6,832 & 6,772  \\
        \hline
        \hline
    \end{tabular}
\end{table}

The convergence behavior is illustrated in Figure~\ref{fig:RMSE_trends}. 
The sinNN models demonstrate a monotonic reduction in error as model complexity ($N$) increases. 
For the largest dataset (75k), the $N=70$ sinNN model achieves a training RMSE of just 17.0~cm$^{-1}$. 
Crucially, the difference between training, test, and validation RMSE is at most 4--5~cm$^{-1}$, confirming that the sinNN models generalize well without overfitting.

\begin{figure}[H]
\centering
    \begin{tikzpicture}
    \begin{axis}[
        width=14cm,
        height=8cm,
        xlabel={Number of neurons},
        ylabel={RMSE (cm$^{-1}$)},
        legend pos=south west,
        xmode=linear,
        ymode=log,
        log y ticks with fixed point,
        ymax=370,
        ymin=0,
        xtick={30,40,50,60,70},
        ytick={0,50,100,150,250,350},
        scaled x ticks=false,  
        tick label style={/pgf/number format/fixed,/pgf/number format/use comma=false,/pgf/number format/set thousands separator={}},
        legend cell align={left},
        legend style={draw=none, font=\normalfont}
    ]
    
    \addplot[color=Purple, mark=square*, thick] coordinates {
        (30,58.2741175957038) (40,39.1206628563597) (50,27.2954835097541) (60,40.6347298792419) (70,23.2008727423658)
    };
    \addlegendentry{25k}

    \addplot[color=BurntOrange, mark=square*, thick] coordinates {
(30,70.8823907233961) (40,50.5988402224168) (50,31.7544982840289) (60,21.4492626594447) (70,29.6304377204643)
    };
    \addlegendentry{50k}
    
    \addplot[color=BlueGreen, mark=square*, thick] coordinates {
(30,82.7914887534553) (40,64.3621457798422) (50,38.1148013522807) (60,35.5412156609653) (70,17.0447934471551)
    };
    \addlegendentry{75k}

    \addplot[dashed, mark= square*, mark options={solid}, color=Purple, thick] coordinates {
(30,173.753382824537) (40,135.459990513999) (50,134.479670061813) (60,142.384044828546) (70,105.392935188691)
    };

        \addplot[dashed, mark= square*, mark options={solid}, color=BurntOrange, thick] coordinates {
(30,145.594440254647) (40,142.774714514099) (50,97.394107677936) (60,83.5644864858199) (70,87.7321792543886)
    };

    \addplot[dashed, mark= square*, mark options={solid}, color=BlueGreen, thick] coordinates {
(30,131.887671634104) (40,109.093041726417) (50,77.4646811026869) (60,70.4764409360867) (70,39.9158908937499)
    };

    \addplot[color=Purple, mark=*, thick] coordinates {
(30,108.203624286604) (40,58.4753643979827) (50,66.9250909820943) (60,55.9268188961093) (70,53.249924211119)
    };

    \addplot[color=BurntOrange, mark=*, thick] coordinates {
(30,108.84837834536) (40,93.3049842603298) (50,70.9818894244662) (60,63.4869662695099) (70,45.5956892261101)
    };
    
    \addplot[color=BlueGreen, mark=*, thick] coordinates {
(30,153.316926712901) (40,82.3404124178679) (50,67.4124483477572) (60,47.0892715076641) (70,55.6245319438988)
    };

    \addplot[dashed, mark= *, mark options={solid}, color=Purple, thick] coordinates {
(30,338.575997296475) (40,287.582400755217) (50,328.202884954168) (60,338.296550383595) (70,300.659979413931)
    };

        \addplot[dashed, mark= *, mark options={solid}, color=BurntOrange, thick] coordinates {
(30,272.398960571202) (40,263.552242095348) (50,178.184694721117) (60,232.606810751488) (70,169.188164909547)
    };

    \addplot[dashed, mark= *, mark options={solid}, color=BlueGreen, thick] coordinates {
(30,223.75961871532) (40,132.781969780527) (50,119.492781297217) (60,84.1478373582895) (70,118.297178476286)
    };

    \end{axis}
\end{tikzpicture}
\caption{Convergence of the Fitting Error (RMSE in cm$^{-1}$) for sinNN ($\blacksquare$) and expNN ($\bullet$) models as a function of neurons and dataset size \cAA{for the refitting of the Richter HONO PES}. Solid lines: Training RMSE; Dashed lines: RMSE with respect to full master dataset.}
\label{fig:RMSE_trends}
\end{figure}

In contrast, the expNN architecture struggles to capitalize on increased data density. 
Even with the 75k dataset, the expNN model stalls at an RMSE of 55.6~cm$^{-1}$. 
Furthermore, expNN models exhibit a persistent generalization gap (\textit{e.g.}, Train 53.2 vs. Valid 87.9~cm$^{-1}$ for 25k/$N$=70). 
This indicates that the unbounded exponential activation functions create a rugged optimization landscape prone to overfitting, whereas the bounded sinusoidal basis offers better stability here.

The most stringent test of the PES quality is the reproduction of anharmonic vibrational eigenvalues. 
Table~\ref{tab:model_comparison} reports the Root Mean Square Deviation (RMSD) of the computed vibrational states relative to the Richter reference. 
The sinNN models achieve spectroscopic accuracy. 
For the 75k dataset, the $N=60$ model reproduces \textit{trans}-HONO energies with an RMSD of only 1.5~cm$^{-1}$. 
However, a bias is evident: the \textit{trans}-isomer is consistently described with higher accuracy than the \textit{cis}-isomer. 
This is an artifact of the sampling strategy, where grids were centered on the \textit{trans}-equilibrium, leading to higher point density in that well. 
Conversely, expNN models fail to reach this precision (best case $\sim$6.0~cm$^{-1}$ for \textit{trans}). 
The difference in both fitting error and vibrational accuracy highlights the suitability of sinusoidal activation functions for compact, high-fidelity PES representations.

\begin{table}[H]
    \centering
    \caption{Comparative performance of sinNN and expNN models \cAA{for the refitting of the Richter HONO PES} across hierarchical datasets and neuron counts ($N$). {RMSDs} with respect to computed vibrational eigenenergies relative to the reference Richter PES (Trans and Cis). RMSEs for static fitting errors on the validation set (Valid), and the error on the full 300,000-point master dataset (300k). All values are in cm$^{-1}$.}
    \label{tab:model_comparison}
    \begin{tabular}{l c | c c c c | c c c c}
        \hline \hline
         & & \multicolumn{4}{c|}{\textbf{sinNN}} & \multicolumn{4}{c}{\textbf{expNN}} \\
        \textbf{Dataset} & \textbf{$N$} & \textbf{Trans} & \textbf{Cis} & \textbf{Valid} & \textbf{300k} & \textbf{Trans} & \textbf{Cis} & \textbf{Valid} & \textbf{300k} \\
        \hline
        25k & 40 & 3.6 & 13.7 & 42.2 & 135.5 & 22.1 & 23.6 & 85.7 & 287.6 \\
        25k & 50 & 4.5 & 14.8 & 36.5 & 134.5 & 31.5 & 21.7 & 99.6 & 328.2 \\
        25k & 60 & 7.6 & 16.4 & 48.7 & 142.4 & 22.2 & 25.9 & 91.6 & 338.3 \\
        25k & 70 & 2.3 & 14.0 & 28.0 & 105.4 & 21.8 & 21.8 & 87.9 & 300.7 \\
        \hline
        50k & 40 & 4.2 & 17.9 & 53.3 & 142.8 & 42.6 & 34.3 & 91.8 & 263.6 \\
        50k & 50 & 4.5 & 10.1 & 34.9 & 97.4 & 21.6 & 30.9 & 69.6 & 178.2 \\
        50k & 60 & 2.5 & 8.8 & 25.1 & 83.6 & 14.7 & 21.3 & 64.1 & 232.6 \\
        50k & 70 & 2.5 & 10.9 & 31.7 & 87.7 & 9.9 & 12.3 & 50.5 & 169.2 \\
        \hline
        75k & 40 & 5.0 & 15.9 & 63.3 & 109.1 & 16.1 & 23.7 & 83.2 & 132.8 \\
        75k & 50 & 1.8 & 11.1 & 38.3 & 77.5 & 13.9 & 13.6 & 68.8 & 119.5 \\
        75k & 60 & \textbf{1.5} & 14.0 & 36.4 & 70.5 & 6.0 & 13.8 & 49.2 & 84.1 \\
        75k & 70 & 3.4 & \textbf{3.7} & \textbf{17.6} & \textbf{39.9} & 14.5 & 16.5 & 57.4 & 118.3 \\
        \hline \hline
    \end{tabular}
\end{table}

\begin{figure}
\centering
    \begin{tikzpicture}
    \begin{axis}[
        width=14cm,
        height=8cm,
        xlabel={Number of neurons},
        ylabel={RMSD (cm$^{-1}$)},
        legend pos=north east,
        xmode=linear,
        ymode=linear,
        ymin=0, ymax=50,
        xtick={30,40,50,60,70},
        ytick={0,10,20,30,40,50},
        minor ytick={5,15,25,35,45},
        scaled x ticks=false,  
        tick label style={/pgf/number format/fixed,/pgf/number format/use comma=false,/pgf/number format/set thousands separator={}},
        legend cell align={left},
        legend style={draw=none, font=\normalfont}
    ]

    \addplot[color=Purple, mark=square*, thick] coordinates {
(30,8.1643) (40,3.6013) (50,4.5188) (60,7.5989) (70,2.3424)
    };
    \addlegendentry{25k}

    \addplot[color=BurntOrange, mark=square*, thick] coordinates {
(30,7.2322) (40,4.1694) (50,4.4546) (60,2.5187) (70,2.5373)
    };
    \addlegendentry{50k}

    \addplot[color=BlueGreen, mark=square*, thick] coordinates {
(30,6.1738) (40,5.049) (50,1.8106) (60,1.5196) (70,3.3924)
    };
    \addlegendentry{75k}

    \addplot[dashed, mark=square*, color=Purple, thick] coordinates {
(30,21.9683) (40,13.6761) (50,14.8133) (60,16.3618) (70,13.9697)
    };

    \addplot[dashed, mark=square*, color=BurntOrange, thick] coordinates {
(30,18.8658) (40,17.8982) (50,10.0587) (60,8.7642) (70,10.8959)
    };

    \addplot[dashed, mark=square*, color=BlueGreen, thick] coordinates {
  (30,15.3633) (40,15.9044) (50,11.0777) (60,13.9549) (70,3.6512)
    };

    \addplot[color=Purple, mark=*, thick] coordinates {
(30,35.826) (40,22.1113) (50,31.5444) (60,22.1809) (70,21.8238)
    };

    \addplot[color=BurntOrange, mark=*, thick] coordinates {
 (30,25.7977) (40,42.5829) (50,21.5905) (60,14.7446) (70,9.8708)
    };

    \addplot[color=BlueGreen, mark=*, thick] coordinates {
 (30,32.9499) (40,16.1421) (50,13.8953) (60,5.996) (70,14.4721)
    };

    \addplot[dashed, mark=*, color=Purple, thick] coordinates {
(30,40.7387) (40,23.5899) (50,21.7205) (60,25.9188) (70,21.7985)
    };

    \addplot[dashed, mark=*, color=BurntOrange, thick] coordinates {
(30,39.6393) (40,34.3001) (50,30.8598) (60,21.2625) (70,12.3382)
    };
    \addplot[dashed, mark=*, color=BlueGreen, thick] coordinates {
(30,46.7482) (40,23.671) (50,13.6333) (60,13.8485) (70,16.4779)
    };

    \end{axis}
    \end{tikzpicture}
\caption{RMSD of fundamental vibrational transition energies of \textit{trans}- (continuous) and \textit{cis}-HONO (dashed) computed from our sinNN ($\blacksquare$) and expNN ($\bullet$) models vs. Richter reference.}
\label{fig:sinNN_vib_Richter}
\end{figure}

\subsubsection{Optimized Sampling: Dual-Reference}
\label{sec:dual-ref}

To eliminate the topological bias observed in the models based on single-reference training datasets, we implemented a dual-reference strategy. 
Two independent sparse grids were generated and merged: one centered at the \textit{trans}-equilibrium and one at the \textit{cis}-equilibrium. 
Two composite datasets were created, a 12.5k+12.5k set (23,326 total points) and a 25k+25k set (47,088 total points).

Table~\ref{tab:dual_ref_results} summarizes the performance of sinNN models on these balanced datasets. 
The results are compelling: the dual-reference strategy successfully symmetrizes the accuracy. 
The best global performance is achieved with the 25k+25k dataset and 80 neurons. 
This model delivers vibrational RMSDs of 2.2~cm$^{-1}$ for \textit{trans}-HONO and 1.8~cm$^{-1}$ for \textit{cis}-HONO. 
Unlike the single-reference approach, where high accuracy in one well came at the cost of the other, the dual-reference model achieves spectroscopic precision ($<$ 2.5~cm$^{-1}$) across the entire relevant configurational space. 
This finding highlights a key strength of the proposed methodology: sparse grid centers can be strategically placed to target physically distinct regions, allowing for systematic convergence even in systems with complex isomerization pathways.

\begin{table}[H]
    \centering
    \caption{Performance of sinNN models trained on dual-reference datasets. RMSD values refer to deviations \cAA{in the HONO} vibrational eigenenergies relative to the Richter benchmark (Trans and Cis). RMSEs for the static fitting errors on the training (Train) and validation (Valid) sets.}
    \label{tab:dual_ref_results}
    \begin{tabular}{l c c c c c}
        \hline \hline
         \textbf{Dataset} & $N$ & \textbf{Trans} & \textbf{Cis} & \textbf{Train} & \textbf{Valid} \\
        \hline
        12.5k+12.5k & 50 & 6.5 & 9.3 & 42.2 & 44.4 \\
        12.5k+12.5k & 60 & 3.2 & 6.2 & 33.1 & 35.3 \\
        12.5k+12.5k & 70 & 3.0 & 5.6 & 24.5 & 27.0 \\
        12.5k+12.5k & 80 & 2.4 & 3.7 & 18.4 & 21.0 \\
        \hline
        25k+25k & 50 & 4.2 & 9.6 & 62.1 & 63.8 \\
        25k+25k & 60 & 4.9 & 6.7 & 31.7 & 34.6 \\
        25k+25k & 70 & 3.7 & 2.7 & 27.1 & 30.5 \\
        25k+25k & 80 & \textbf{2.2} & \textbf{1.8} & \textbf{16.1} & \textbf{17.9} \\
        \hline \hline
    \end{tabular}
\end{table}

\cAA{To contextualize these results, we compare our methodology with previously reported SOP PESs for the ground electronic state of HONO. The current state-of-the-art is the model by Richter \textit{et al.},\cite{ric04:1306} which utilized adaptive sampling and physics-inspired basis functions. Their model, constructed from only 638 carefully chosen \textit{ab initio} points, reproduces experimental fundamental transitions with an RMSD of 6.2 and 4.6 cm$^{-1}$ for \textit{cis}- and \textit{trans}-HONO, respectively. Pradhan and Brown\cite{pccp_19_22272} later reported an expNN PES model using 11,000 points obtained through a combination of low-dimensional grids (limited to rank 2) and random sampling with a predictive energy filter. Their model achieved RMSDs of 8.2 and 9.7 cm$^{-1}$ against experiment. Aerts \textit{et al.}\cite{aerts2022} reached similar precision using comparable electronic structure methods, suggesting that the bottleneck for experimental agreement here lies in the underlying \textit{ab initio} theory rather than the fitting form.}

\cAA{In comparison, while our sinNN models achieve higher precision relative to the underlying benchmark ($<$ 2.5 cm$^{-1}$), our economy of points is roughly worse by a factor of two compared to the work of Pradhan and Brown. We attribute this to our reliance on a rigid, unbiased sparse grid which, while offering a clear path to systematic improvement, does not incorporate a density distribution to favor low-energy regions. Furthermore, we note the work of Avila and Carrington,\cite{avi15:044106} who demonstrated that Smolyak interpolation could reproduce the Richter PES with high accuracy. They reported a remarkable economy of points requiring as few as 1,573 points when using non-ideal basis functions. However, this efficiency was achieved through coordinate bounds optimization. In contrast, our current approach prioritizes an automated fitting procedure that does not require such system-specific coordinate manipulations or \textit{a priori} knowledge of the surface topology.}

\cAA{It is also instructive to compare this with the low-rank fitting methods of Peláez and coworkers.\cite{pel13:014108,pan20:234110,nad23:114109} Their 2013 MGPF implementation for HONO relied on 1.15 million points, which was subsequently reduced to approximately 37,500 points using SOP-FBR and 52,500 points with CP-FBR. While these methods achieve high operator compactness, a direct comparison of model quality is complicated by differences in the fitting objectives. Specifically, these previous models employ an energy cutoff significantly higher than the recommendation for the original Richter surface, and the reported vibrational transition energies are computed on a relatively coarse grid.}

\cAA{To address the point economy within our framework, we investigated limiting the rank of the subgrids to 3 while maintaining independent dual-reference grids. This effectively reduces the 12.5k+12.5k dataset to 16,786 points. A 70-neuron sinNN model trained on this reduced set achieves a validation RMSE of 33.4 cm$^{-1}$ and vibrational RMSDs of 4.3 cm$^{-1}$ for both isomers. This result indicates that strategically truncating higher-order couplings can significantly improve data efficiency without sacrificing the spectroscopic accuracy required for quantum dynamics simulations. This needs to be further explored and generalized.}

\subsection{AIQM2-based PES Construction and Validation}
\label{sec:aiqm2_pes}

The second phase of our evaluation involves the direct fitting of PES models from potential energies computed using the AI-enhanced method MLatom/AIQM2.\cite{mlatom3,Chen2024-qa} 
There is currently strong interest in quantifying AIQM2's ability to support anharmonic vibrational spectroscopy. 
Here, we evaluate this by comparing computed vibrational transition frequencies against available high-resolution experimental data.

\subsubsection{HONO: Spectroscopic Accuracy vs. Experiment}
\label{sec:aiqm2_hono}

We constructed a PES for HONO using AIQM2 energies and the optimized dual-reference sampling strategy (12.5k \textit{trans} + 12.5k \textit{cis}) \cAA{with highest grid rank equal to 4}. 
After energy filtering, the dataset contained 23,384 points. 
sinNN models with 50 to 80 neurons were trained, with results summarized in Table~\ref{tab:aiqm2_results}.

\begin{table}[H]
    \centering
    \caption{Vibrational fundamental frequencies (cm$^{-1}$) of HONO computed from AIQM2-based sinNN models compared to experiment.\cite{dee83:199} RMSD Exp: Deviation from experiment. RMSE: Static fitting error.}
    \label{tab:aiqm2_results}
    \begin{tabular}{l | c c | c c | c c | c c| c c}
        \hline \hline
        $N$ & \multicolumn{2}{c|}{\textbf{50}} &\multicolumn{2}{c|}{\textbf{60}} & \multicolumn{2}{c|}{\textbf{70}} & \multicolumn{2}{c|}{\textbf{80}} & \multicolumn{2}{c}{\textbf{Experiment}} \\
        \textbf{Mode} & \textit{cis} & \textit{trans} & \textit{cis} & \textit{trans}& \textit{cis} & \textit{trans} & \textit{cis} & \textit{trans} & \textit{cis} & \textit{trans} \\
        \hline
        $\nu_{\ce{N=O}}$ & 1672.9 & 1729.3 & 1671.2 & 1731.5 & 1670.3 & 1729.1 & 1669.8 & 1727.7 & 1640.5 & 1699.8 \\
        $\nu_{\ce{NO}}$ & 862.4 & 812.9 & 868.5 & 810.5 & 864.5 & 810.6 & 864.7 & 812.6 & 851.9 & 790.1 \\
        $\nu_{\ce{OH}}$ & 3427.9 & 3592.8 & 3421.5 & 3595.4 & 3422.3 & 3594.7 & 3423.0 & 3595.9 & 3426.2 & 3590.8 \\
        $\delta_{\ce{ONO}}$ & 622.9 & 607.1 & 623.5 & 607.8 & 624.0 & 605.8 & 620.3 & 608.2  & 609.2 & 595.6 \\
        $\delta_{\ce{HON}}$ & 1310.1 & 1277.1 & 1315.5 & 1278.6 & 1316.2 & 1278.5 & 1314.0 & 1275.5 & 1313.1 & 1263.2 \\
        $\tau$ & 638.3 & 529.1 & 640.7 & 534.6 & 646.5 & 527.7 & 640.2 & 533.7 & 639.7 & 543.9 \\
        \hline
        \textbf{RMSD Exp} & \textbf{15.1} & \textbf{18.0} & \textbf{15.6} & \textbf{17.9} & \textbf{14.9} & \textbf{17.8} & \textbf{13.9} & \textbf{17.0} & -- & -- \\
        \hline
        {RMSE Train} & \multicolumn{2}{c|}{41.0} & \multicolumn{2}{c|}{33.2} & \multicolumn{2}{c|}{23.2} & \multicolumn{2}{c|}{23.1} & \multicolumn{2}{c}{--} \\
        \hline \hline
    \end{tabular}
\end{table}

The static fitting error follows expected convergence trends, dropping from 41.0 to 23.1~cm$^{-1}$ as the network grows to 80 neurons. 
However, the RMSD with respect to experimental vibrational frequencies plateaus at $\sim$14--17~cm$^{-1}$. 
This plateau indicates that the sinNN model should have converged to the underlying AIQM2 landscape; the remaining deviation represents the intrinsic accuracy limit of the AIQM2 method itself, rather than a fitting artifact.

Overall, the ability to obtain a PES with a global vibrational RMSD of roughly 16~cm$^{-1}$ using only $\sim$23,000 single-point calculations demonstrates the reasonable data efficiency of the combined sinNN/sparse-grid approach for predictive spectroscopy. 
Moreover, RMSDs with respect to experimental data fall within the same order of magnitude with the ones expected from PES models trained on potential energies obtained from \textit{ab initio} methods that are thousands of times more computationally expensive.\cite{aerts2022} 

\cAA{Regarding computational efficiency, the overhead of the sinNN fitting procedure is negligible compared to the time required for subsequent quantum dynamics calculations. For the systems considered, potential energy evaluation via AIQM2 requires less than 3 seconds per point. In contrast, obtaining a single potential energy at the CCSD(T)-F12/cc-pVTZ-F12 level using MOLPRO\cite{MOLPRO_long} takes approximately 6.5 CPU-minutes on the same hardware. While the evaluation of any PES on the massive grids required for MCTDH simulations entails a non-negligible computational cost, the compact SOP form of the sinNN models ensures that this process is highly efficient. This minimizes the overhead of the PES representation, preventing it from becoming a disproportionate bottleneck during wavepacket propagation.}

\subsubsection{Assessing Robustness: Formic and Carbamic Acid}
\label{subsec:larger_systems}


The calculation of vibrational eigenstates via any wavefunction-based method requires a PES that is physically meaningful and globally well-behaved across the entire coordinate range. Variational solutions exhibit a pathological sensitivity to ``spurious holes'' or unphysical depressions; even a minor unphysical minimum can act as an artificial global sink. This causes the variational algorithm to collapse the wavefunction into the artifact to minimize energy, resulting in a catastrophic failure of the bound-state solver and a violation of the physical assumptions of the calculation.

To validate the robustness of the sinNN/AIQM2 approach for higher-dimensional problems, we first examined formic acid (\ce{HCOOH}), a prototypical organic acid with two stable conformers (\textit{cis} and \textit{trans}). 
Verification of the equilibrium geometries obtained from AIQM2, using Gaussian\cite{g16} as the optimization engine, against a high-level CCSD(T) reference\cite{puz24:2501} (see Supplementary Material) confirms that AIQM2 accurately captures subtle structural features.

Building on these results, we extended the analysis to carbamic acid (\ce{H2NCOOH}), a prebiotic molecule of significant interest for interstellar detection.\cite{puz24:2501} Carbamic acid presents a more complex topological challenge due to the non-planarity of its \ce{NH_2} group. 
As shown in the Supplementary Material, AIQM2 accurately reproduces the non-planar equilibrium structure, with the \ce{NH_2} group deviating from the molecular plane by approximately 15$^\circ$ ($\tau_{\text{H6}} \approx 166^\circ$).

Importantly, we verified the nature of these stationary points by computing the harmonic frequencies at the AIQM2 level. The planar configuration was confirmed as a transition state, characterized by a single imaginary frequency corresponding to the \ce{NH_2} inversion mode. 

Global PES models were constructed using 80-neuron sinNN architectures trained on hierarchical sparse grids of rank up to 4 (80,000 points for \ce{HCOOH} and 100,000 for \ce{H2NCOOH}). The potential energies for \ce{HCOOH} were obtained using the dual-reference sampling strategy discussed previously. As summarized in Table~\ref{tab:nn_performance}, the fitting performance remained consistent, with RMSE values below 100~cm$^{-1}$ despite the increased dimensionality, and extended energy range as compared to the HONO benchmark. 

The final and most rigorous validation was provided by variational calculations of the ground vibrational states. The calculations yielded stable, physically sound Zero-Point Energies (ZPE) for all studied systems, which are given in Table~\ref{tab:gs_energies} \cAA{for completeness}. 
The absence of wavefunction collapse into unphysical regions confirms that the sinNN/sparse-grid framework produces a topologically correct landscape, free of spurious artifacts. These results demonstrate that the combination of systematic hierarchical sampling and sinusoidal activation functions provides a reliable and automated route to high-fidelity Potential Energy Surfaces for this class of molecules.

\begin{table}[H]
\centering
\caption{Neural network static fitting performance (RMSE) and calculated ZPEs. All reported values in cm\textsuperscript{-1}.}
\label{tab:nn_performance}
\begin{tabular}{l|cc|cc}
\hline \hline
  & \multicolumn{2}{c|}{\ce{HCOOH}}  & \multicolumn{2}{c}{\ce{H2NCOOH}}\\ 
 &Train&Valid&Train&Valid\\ \hline
RMSE &  37.7    & 40.6 & 117    &123\\
\hline 
ZPE & \multicolumn{2}{c|}{\textit{cis}: 9059 / \textit{trans}: 7569} & \multicolumn{2}{c}{11193} \\
\hline \hline
\end{tabular}
\label{tab:gs_energies}
\end{table}

\section{Conclusions and perspectives}
\label{sec:concl}

\cAA{The construction of accurate, global PESs in SOP form remains a critical bottleneck for high-dimensional quantum dynamics simulations. In this work, we have presented an automated methodology that combines hierarchical sparse grid sampling with sinusoidal neural networks (sinNN) to address this challenge. The resulting potential models are in the SOP form required for the efficiency of the Multi-Configuration Time-Dependent Hartree (MCTDH) method, remaining compact and physically robust within the configuration space of interest. While sinNN models should be considered a first resort, as they may not compete with specialized methods in terms of point economy or minimal expression size, this framework provides an automated tool interfaced with the Heidelberg implementation of MCTDH.}

\cAA{A central finding of this study is that periodic activation functions in the sinNN architecture offer superior numerical stability compared to traditional exponential-based networks (expNN) for the applications considered in this work. The sinNN models demonstrate improved generalization and avoid the overfitting characteristic of unbounded functions, all while maintaining the strict SOP form. Furthermore, the introduction of a Dual-Reference sampling strategy proved decisive for capturing complex molecular topologies, such as the double-well potential of HONO. By fusing isomer-centered grids, we successfully symmetrized the accuracy for HONO, yielding a global PES that reproduces vibrational eigenstates with spectroscopic precision ($\sim$2.5 cm$^{-1}$). The application to AIQM2 potential energies further confirms the accessibility of the method, demonstrating that this approach is capable of describing PES landscapes far beyond equilibrium geometries. This allows for the generation of topologically sound global surfaces for medium-sized molecules with reasonable data efficiency.}

\cAA{Looking forward, we aim to take advantage of the nested nature of the grids and the interpolative approach outlined in Section~\ref{sec:energy_filter} to develop an inherently adaptive sampling scheme. Unlike traditional active learning loops in machine learning, the hierarchical coefficients, or ``surpluses,'' provide a direct measure of refinement fully adapted to the interpolation. Moreover, because the subgrids directly sample the terms of a High-Dimensional Model Representation (HDMR), following strict hierarchical ancestry of the points ensures that the lower-order terms of the expansion are exactly known at the coordinates of higher-dimensional points. This nesting property avoids the convergence issues common in standard HDMR expansions, where instabilities arise if the low-order background is not sufficiently resolved before higher-order couplings are introduced. By coupling adaptive sampling driven by hierarchical refinement with an interpolative scheme intrinsically tailored to the sampled grid, the point economy for constructing accurate SOP PESs may be significantly enhanced through a systematically improvable and adaptive framework. Although, the models are unlikely to be as compact.}

\cAA{Finally, we identify a clear path toward balancing high-level accuracy with the demands of high dimensionality. State-of-the-art PES fitting approaches leverage the nearsightedness of atomic interactions and molecular similarity, using input functions that encode the atoms local chemical environments.\cite{beh11:074106,beh21:10037} These methods are currently incompatible with the strict SOP requirement and may not be representable in such a form. The approach proposed here thus provides a useful framework, where a highly effective method like AIQM2 is used to obtain a baseline model from a large number of points, kept compact by sinNN fitting. This baseline can then be corrected by another model built from relatively small number of high-level \textit{ab initio} energies. By combining the data efficiency of an adaptive interpolative scheme with the compactness of sinNN fitting, we provide a robust and automated route toward high-fidelity PES generation for complex molecular systems within the MCTDH framework.}

\section*{Supplementary material}

The supplementary material accompanying this manuscript includes: the code to run expNN and sinNN fitting, 
detailed improved relaxation parameters used in the MCTDH computations, 
a complete set of input files for running the improved relaxation computations, the MCTDH operators generated for all trained models, and computed vibrational transition energies for both \textit{trans}- and \textit{cis}-HONO isomers from all trained models.

\section*{Acknowledgments}
The author acknowledges the Institut Interuniversitaire des Sciences Nucléaires (IISN) Grant No. 4.4504.10 and FRS-FNRS for Postdoctoral Researcher Grant No. 40032227 for financial support. 
I warmly thank Dr. Daniel Hurtmans (ULB) for useful discussions and Prof. Henrik R. Larsson (UC Merced) for pointing out the trigonometric identity used to factorize the sinusoidal basis functions.

\section*{Data}
The data that support the findings of this study are available from the corresponding author upon reasonable request. 

\section*{Disclosure statement}
The authors declares no conflict of interest.

\bibliographystyle{apsrev4-1}

\bibliography{refs}

@string{arpc={Ann.\ Rev.\ Phys.\ Chem.}}

@string{ap={Ann.\ Phys.}}

@string{cp={Chem.\ Phys.}}

@string{cpl={Chem.\ Phys.\ Lett.}}

@string{crv={Chem.\ Rev.}}

@string{cs={Chemical Science}}

@string{jcc={J.~Comp.\ Chem.}}

@string{jcp={J.~Chem.\ Phys.}}

@string{jctc={J.~Chem.\ Theory  Comput.}}

@string{jiam={J.\ Soc.\ Indust.\ Appl.\ Math.}}

@string{jms={J.~Mol.\ Struct.\ (Theochem)}}

@string{jmsp={J.~Mol.\ Spec.}}

@string{jpca={J.~Phys.\ Chem.~A}}

@string{jtcc={J.~Theor.\ Comp.\ Chem.}}

@string{ma={Math.\ Ann.}}

@string{mp={Mol.\ Phys.}}

@string{natc={Nat.\ Commun.}}

@string{pccp={Phys.\ Chem.\ Chem.\ Phys.}}

@string{prep={Phys.\ Rep.}}

@string{tca={Theor.\ Chim.\ Acta}}

@string{TCA={Theor.\ Chem.\ Acc.}}

@article{bec00:1,
author={M. H. Beck and A. J{\"a}ckle and G. A. Worth and H.-D. Meyer},
title={The multi-configuration time-dependent {H}artree ({MCTDH})
method: {A} highly efficient algorithm for propagating wave packets},
journal={Phys.~Rep.},year={2000},volume={324},pages={1-105}}

@article{bor27:457,
author={M. Born and R. Oppenheimer},
title={Zur {Q}uantentheorie der {M}olekeln},
journal=ap,volume={84},year={1927},pages={457}}

@article{bra09:577,
author={B. J. Braams and J. M. Bowman},
title={Permutationally invariant potential energy surfaces in
       high dimensionality},
journal={Int.~Rev.~Phys.~Chem.},year={2009},volume={28},pages={577-606},
doi={10.1080/01442350903234923}}

@article{dee83:199,
author={C. M. Deely and I. M. Mills},
title={The infrared vibration-rotation spectrum
   of {\it trans} and {\it cis} nitrous acid},
journal=jms,year={1983},volume={100},pages={199}}

@article{dor08:224109,
author={L. J. Doriol and F. Gatti and C. Iung and H.-D. Meyer},
title={Computation of vibrational energy levels and eigenstates
   of fluoroform using the multiconfiguration time-dependent
   {H}artree method},
journal=jcp,year={2008},volume={129},pages={224109}}

@article{gat09:1,
author={F. Gatti and C. Iung},
title={Exact and constrained kinetic energy operators for polyatomic
       molecules: The polyspherical approach},
journal=prep,year={2009},volume={484},pages={1-69}}

@article{koc14:021101,
author={W. Koch and D. H. Zhang},
title={Communication: Separable potential energy surfaces from
       multiplicative artificial neural networks},
journal=jcp,year={2014},volume={141},pages={021101}}

@misc{MOLPRO_long,
title={MOLPRO, version 2025.3, a package of ab initio programs},
author={
H.-J. Werner and
P. J. Knowles and

P. Celani and
W. Gy\"orffy and
A. Hesselmann and
D. Kats and
G. Knizia and
A. K\"ohn and
T. Korona and
D. Kreplin and
R. Lindh and
Q. Ma and
F. R. Manby and
A. Mitrushenkov and
G. Rauhut and
M. {Sch\"{u}tz} and
K. R. Shamasundar and

T. B. Adler and
R. D. Amos and
S. J. Bennie and
A. Bernhardsson and
A. Berning and
J. A. Black and
P. J. Bygrave and
R. Cimiraglia and
D. L. Cooper and
D. Coughtrie and
M. J. O. Deegan and
A. J. Dobbyn and
K. Doll and 
M. Dornbach and
F. Eckert and
S. Erfort and
E. Goll and
C. Hampel and
G. Hetzer and
J. G. Hill and
M. Hodges and 
T. Hrenar and
G. Jansen and
C. K\"oppl and
C. Kollmar and
S. J. R. Lee and
Y. Liu and
A. W. Lloyd and
R. A. Mata and
A. J. May and
B. Mussard and
S. J. McNicholas and
W. Meyer and
T. F. {Miller III} and
M. E. Mura and
A. Nicklass and
D. P. O'Neill and
P. Palmieri and
D. Peng and
K. A. Peterson and
K. Pfl\"uger and
R. Pitzer and
I. Polyak and
M. Reiher and
J. O. Richardson and
J. B. Robinson and
B. Schr\"oder and
M. Schwilk and 
T. Shiozaki and
M. Sibaev and
H. Stoll and
A. J. Stone and
R. Tarroni and
T. Thorsteinsson and
J. Toulouse and
M. Wang and
M. Welborn and 
B. Ziegler
},
note={see http://www.molpro.net},
address={Stuttgart, Germany},year={2025}}

@article{man92:3199,
title={Wave-Packet Dynamics within the Multiconfiguration {H}artree
Framework: General Aspects and application to {NOCl}},
author={U. Manthe and H.-D. Meyer and L. S. Cederbaum},
journal=jcp,year={1992},volume={97},pages={3199-3213}}

@ARTICLE{man06:194105,
author={S. Manzhos and T. {Carrington, Jr.}},
title={Using neural networks to represent potential surfaces
       as sum of products},
journal=jcp,year={2006},volume={125},pages={194105}}

@ARTICLE{man07:014103,
author={S. Manzhos and T. {Carrington, Jr.}},
title={Using redundant coordinates to represent potential energy
       surfaces with lower-dimensional functions},
journal=jcp,year={2007},volume={125},pages={014103}}

@article{mar63:431,
author={D. Marquardt},
title={An algorithm for least squares estimation of non-linear parameters},
journal=jiam,volume={11},year={1963},pages={431 - 441}}

@misc{mctdh:package,
author={G. A. Worth and M. H. Beck and A. J{\"a}ckle and
    H.-D. Meyer},
howpublished={The {MCTDH} {P}ackage,
          H.-D. Meyer,          Used version: 8.6.
          {S}ee http://mctdh.uni-hd.de/ for a description
           of the Heidelberg MCTDH package.},
organization={University of Heidelberg, Germany}}

@article{mey90:73,
title={The Multi-Configurational Time-Dependent {H}artree Approach},
author={H.-D. Meyer and U. Manthe and L. S. Cederbaum},
journal=cpl,year={1990},volume={165},pages={73-78}}

@article{mey03:251,
author={H.-D. Meyer and G. A. Worth},
title={{Q}uantum molecular dynamics:
   {P}ropagating wavepackets and density operators using the
   multiconfiguration time-dependent {H}artree ({MCTDH}) method},
journal=TCA,year={2003},volume={109},pages={251-267}}

@article{mey06:179,
author={H.-D. Meyer and F. {Le Qu\'er\'e} and C. L\'eonard
    and F. Gatti},
title={Calculation and selective population of vibrational levels with
the {M}ulticonfiguration {T}ime-{D}ependent {H}artree ({MCTDH})
algorithm},
journal=cp,year={2006},volume={329},pages={179-192}}

@article{ndo12:034107,
author={M. Ndong and L. {Joubert Doriol} and H.-D. Meyer
        and A. Nauts and F. Gatti and D. Lauvergnat},
title={Automatic computer procedure for generating exact and analytical
       kinetic energy operators based on the polyspherical approach},
journal=jcp,year={2012},volume={136},pages={034107}}

@article{ndo13:204107,
author={M. Ndong and A. Nauts and L. {Joubert-Doriol} and H.-D. Meyer
        and F. Gatti and D. Lauvergnat},
title={Automatic computer procedure for generating exact and analytical
       kinetic energy operators based on the polyspherical approach:
       general formulation and removal of singularities},
journal=jcp,year={2013},volume={139},pages={204107},
doi={10.1063/1.4828729}}

@article{pel13:014108,
author={D. Pel{\'a}ez and H.-D. Meyer},
title={{The multigrid POTFIT (MGPF) method: Grid representations
        of potentials for quantum dynamics of large systems}},
journal=jcp,year={2013},volume={138},pages={014108}}

@article{pra16:158,
author={E. Pradhan and A. Brown},
title={Neural Network Exponential Fitting of a Potential Energy
       Surface with Multiple Minima: {A}pplication to {HFCO}},
journal=jmsp,year={2016},volume={330},pages={158-164}}

@article{pra16:174305,
author={E. Pradhan and A. Brown},
title={Vibrational energies for {HFCO} using a neural network sum
       of exponentials potential energy surface},
journal=jcp,year={2016},volume={144},pages={174305}}

@article{ric04:1306,
author={F. Richter and M. Hochlaf and P. Rosmus and F. Gatti and H.-D. Meyer},
title={A study of mode--selective trans--cis isomerisation
   in {HONO} using ab initio methodology},
journal=jcp,volume={120},year={2004},pages={1306--1317}}

@article{ric14:1401,
author={F. Richter and P. Carbonniere and C. Pouchan},
title={Towards linear scaling: {L}ocality of potential energy surface
       coupling in valence coordinates},
journal=jcp,year={2014},volume={114},pages={1401}}

@article{wan15:7951,
author={H. Wang},
title={{Multilayer Multiconfiguration Time-Dependent Hartree Theory}},
journal=jpca,year={2015},volume={119},pages={7951}}

@article{pccp_26_1829,
  title={Benchmarking non-adiabatic quantum dynamics using the molecular Tully models},
  author={G{\'o}mez, Sandra and Spinlove, Eryn and Worth, Graham},
  journal=pccp,
  volume={26},
  number={3},
  pages={1829--1844},
  year={2024},
  publisher={Royal Society of Chemistry}
}

@article{wircms_12_e1605,
  title={Vibrational spectroscopy by means of first-principles molecular dynamics simulations},
  author={Ditler, Edward and Luber, Sandra},
  journal={Wiley Interdiscip. Rev. Comput. Mol. Sci.},
  volume={12},
  number={5},
  pages={e1605},
  year={2022},
  publisher={Wiley Online Library}
}

@book{book_bowman,
author = {Bowman, Joel M},
title = {Vibrational Dynamics of Molecules},
publisher = {WORLD SCIENTIFIC},
year = {2022},
address = {},
edition   = {},
doi={10.1142/12305}
}

@Article{Chen2024-qa,
author ="Chen, Yuxinxin and Dral, Pavlo O.",
title  ="AIQM2: organic reaction simulations beyond DFT",
journal  ="Chem. Sci.",
year  ="2025",
volume  ="16",
issue  ="35",
pages  ="15901-15912",
publisher  ="The Royal Society of Chemistry"}

@article{franco2024pyrogenic,
  title={Pyrogenic {HONO} seen from space: insights from global {IASI} observations},
  author={Franco, Bruno and Clarisse, Lieven and Theys, Nicolas and Hadji-Lazaro, Juliette and Clerbaux, Cathy and Coheur, Pierre},
  journal={Atmos. Chem. Phys.},
  volume={24},
  number={8},
  pages={4973--5007},
  year={2024},
  publisher={Copernicus Publications G{\"o}ttingen, Germany}
}

@article{cybenko1989approximation,
  title={Approximation by superpositions of a sigmoidal function},
  author={Cybenko, George},
  journal={Math. Control Signals Syst.},
  volume={2},
  number={4},
  pages={303--314},
  year={1989},
  publisher={Springer}
}

@article{narkhede2022review,
  title={A review on weight initialization strategies for neural networks},
  author={Narkhede, Meenal V and Bartakke, Prashant P and Sutaone, Mukul S},
  journal={Artif. Intell. Rev.},
  volume={55},
  number={1},
  pages={291--322},
  year={2022},
  publisher={Springer}
}

@inproceedings{nguyen1990improving,
  title={Improving the learning speed of 2-layer neural networks by choosing initial values of the adaptive weights},
  author={Nguyen, Derrick and Widrow, Bernard},
  booktitle={1990 IJCNN international joint conference on neural networks},
  pages={21--26},
  year={1990},
  organization={IEEE}
}

@article{bungartz2004sparse,
  title={Sparse grids},
  author={Bungartz, Hans-Joachim and Griebel, Michael},
  journal={Acta Numer.},
  volume={13},
  pages={147--269},
  year={2004},
  publisher={Cambridge University Press}
}

@article{jctc_15_1652,
  title={{GFN2-xTB}—An accurate and broadly parametrized self-consistent tight-binding quantum chemical method with multipole electrostatics and density-dependent dispersion contributions},
  author={Bannwarth, Christoph and Ehlert, Sebastian and Grimme, Stefan},
  journal=jctc,
  volume={15},
  number={3},
  pages={1652--1671},
  year={2019},
  publisher={ACS Publications}
}

@article{smith2017ani,
  title={{ANI}-1: an extensible neural network potential with {DFT} accuracy at force field computational cost},
  author={Smith, Justin S and Isayev, Olexandr and Roitberg, Adrian E},
  journal=cs,
  volume={8},
  number={4},
  pages={3192--3203},
  year={2017},
  publisher={Royal Society of Chemistry}
}

@article{gao2020torchani,
  title={Torch{ANI}: a free and open source {P}y{T}orch-based deep learning implementation of the {ANI} neural network potentials},
  author={Gao, Xiang and Ramezanghorbani, Farhad and Isayev, Olexandr and Smith, Justin S and Roitberg, Adrian E},
  journal={J. Chem. Inf. Model.},
  volume={60},
  number={7},
  pages={3408--3415},
  year={2020},
  publisher={ACS Publications}
}

@article{natc_12_7022,
  title={Artificial intelligence-enhanced quantum chemical method with broad applicability},
  author={Zheng, Peikun and Zubatyuk, Roman and Wu, Wei and Isayev, Olexandr and Dral, Pavlo O},
  journal=natc,
  volume={12},
  number={1},
  pages={7022},
  year={2021},
  publisher={Nature Publishing Group UK London}
}

@article{pccp_26_11469,
  title={Efficient vibrationally correlated calculations using n-mode expansion-based kinetic energy operators},
  author={Bader, Frederik and Lauvergnat, David and Christiansen, Ove},
  journal=pccp,
  volume={26},
  number={15},
  pages={11469--11481},
  year={2024},
  publisher={Royal Society of Chemistry}
}

@article{jcp_136_224105,
  title={An adaptive potential energy surface generation method using curvilinear valence coordinates},
  author={Richter, Falk and Carbonni{\`e}re, Philippe and Dargelos, Alain and Pouchan, Claude},
  journal=jcp,
  volume={136},
  number={22},
  year={2012},
  pages={224105},
  publisher={AIP Publishing}
}

@misc{g16,
author={M. J. Frisch and G. W. Trucks and H. B. Schlegel and G. E. Scuseria and M. A. Robb and J. R. Cheeseman and G. Scalmani and V. Barone and G. A. Petersson and H. Nakatsuji and X. Li and M. Caricato and A. V. Marenich and J. Bloino and B. G. Janesko and R. Gomperts and B. Mennucci and H. P. Hratchian and J. V. Ortiz and A. F. Izmaylov and J. L. Sonnenberg and D. Williams-Young and F. Ding and F. Lipparini and F. Egidi and J. Goings and B. Peng and A. Petrone and T. Henderson and D. Ranasinghe and V. G. Zakrzewski and J. Gao and N. Rega and G. Zheng and W. Liang and M. Hada and M. Ehara and K. Toyota and R. Fukuda and J. Hasegawa and M. Ishida and T. Nakajima and Y. Honda and O. Kitao and H. Nakai and T. Vreven and K. Throssell and Montgomery, {Jr.}, J. A. and J. E. Peralta and F. Ogliaro and M. J. Bearpark and J. J. Heyd and E. N. Brothers and K. N. Kudin and V. N. Staroverov and T. A. Keith and R. Kobayashi and J. Normand and K. Raghavachari and A. P. Rendell and J. C. Burant and S. S. Iyengar and J. Tomasi and M. Cossi and J. M. Millam and M. Klene and C. Adamo and R. Cammi and J. W. Ochterski and R. L. Martin and K. Morokuma and O. Farkas and J. B. Foresman and D. J. Fox},
title={Gaussian~16 {R}evision {C}.01},
year={2016},
note={{G}aussian Inc. Wallingford CT}
}

@article{mlatom3,
	author = {Dral, Pavlo O. and Ge, Fuchun and Hou, Yi-Fan and Zheng, Peikun and Chen, Yuxinxin and Barbatti, Mario and Isayev, Olexandr and Wang, Cheng and Xue, Bao-Xin and Pinheiro Jr, Max and Su, Yuming and Dai, Yiheng and Chen, Yangtao and Zhang, Lina and Zhang, Shuang and Ullah, Arif and Zhang, Quanhao and Ou, Yanchi},
	date = {2024/02/13},
	journal = jctc,
	month = {02},
	number = {3},
	pages = {1193--1213},
	publisher = {American Chemical Society},
	title = {{MLatom} 3: A Platform for Machine Learning-Enhanced Computational Chemistry Simulations and Workflows},
	volume = {20},
	year = {2024},
	year1 = {2024}}

@article{aerts2022,
  title={Adaptive fitting of potential energy surfaces of small to medium-sized molecules in sum-of-product form: Application to vibrational spectroscopy},
  author={Aerts, Antoine and Sch{\"a}fer, Moritz R and Brown, Alex},
  journal={J. Chem. Phys.},
  volume={156},
  number={16},
  pages={164106},
  year={2022},
  publisher={AIP Publishing LLC}
}

@article{pccp_19_22272,
  title={A ground state potential energy surface for {HONO} based on a neural network with exponential fitting functions},
  author={Pradhan, Ekadashi and Brown, Alex},
  journal=pccp,
  volume={19},
  number={33},
  pages={22272--22281},
  year={2017},
  publisher={Royal Society of Chemistry}
}

@article{moh05:65,
  title={Trigonometric identities and sums of separable functions},
  author={Mohlenkamp, Martin J and Monz{\'o}n, Lucas},
  journal={Math. Intell.},
  volume={27},
  number={2},
  pages={65--69},
  year={2005}
}

@incollection{yu22:296,
  title={MULTIMODE, the n-mode Representation of the Potential and Illustrations to IR Spectra of Glycine and Two Protonated Water Clusters},
  author={Yu, Qi and Qu, Chen and Houston, Paul L and Conte, Riccardo and Nandi, Apurba and Bowman, Joel M},
  booktitle={Vibrational Dynamics of Molecules},
  pages={296--339},
  year={2022},
  publisher={World Scientific}
}

@article{puz24:2501,
  title={Carbamic acid and its dimer: A computational study},
  author={Puzzarini, Cristina and Alessandrini, Silvia},
  journal=jcc,
  volume={45},
  number={29},
  pages={2501--2512},
  year={2024},
  publisher={Wiley Online Library}
}

@article{ric18:064303,
  title={Vibrational treatment of the formic acid double minimum case in valence coordinates},
  author={Richter, Falk and Carbonni{\`e}re, Philippe},
  journal=jcp,
  volume={148},
    pages={064303},
  number={6},
  year={2018},
  publisher={AIP Publishing}
}

@article{man08:224104,
  title={Using neural networks, optimized coordinates, and high-dimensional model representations to obtain a vinyl bromide potential surface},
  author={Manzhos, Sergei and Carrington, Tucker},
  journal=jcp,
  volume={129},
  number={22},
  pages={224104},
  year={2008},
  publisher={AIP Publishing}
}

@incollection{MANZHOS2023355,
title = {Chapter 16 - Machine learning for vibrational spectroscopy},
editor = {Pavlo O. Dral},
booktitle = {Quantum Chemistry in the Age of Machine Learning},
publisher = {Elsevier},
pages = {355-390},
year = {2023},
isbn = {978-0-323-90049-2},
doi = {https://doi.org/10.1016/B978-0-323-90049-2.00027-5},
url = {https://www.sciencedirect.com/science/article/pii/B9780323900492000275},
author = {Sergei Manzhos and Manabu Ihara and Tucker Carrington}
}

@article{pra20:e1674936,
author = {Ekadashi Pradhan and Alex Brown},
title = {The lowest lying excited electronic states for HFCO including a potential energy surface for S1 in sum-of-products form},
journal = mp,
volume = {118},
number = {12},
pages = {e1674936},
year = {2020},
publisher = {Taylor \& Francis},
doi = {10.1080/00268976.2019.1674936}
}

@article{avi15:044106,
  title={Using multi-dimensional Smolyak interpolation to make a sum-of-products potential},
  author={Avila, Gustavo and Carrington, Tucker},
  journal=jcp,
  volume={143},
  number={4},
  year={2015},
  pages = {044106},
  publisher={AIP Publishing}
}

@article{man20:10187,
  title={Neural network potential energy surfaces for small molecules and reactions},
  author={Manzhos, Sergei and Carrington Jr, Tucker},
  journal=crv,
  volume={121},
  number={16},
  pages={10187--10217},
  year={2020},
  publisher={ACS Publications}
}

@article{beh21:10037,
  title={Four generations of high-dimensional neural network potentials},
  author={Behler, Jorg},
  journal=crv,
  volume={121},
  number={16},
  pages={10037--10072},
  year={2021},
  publisher={ACS Publications}
}

@article{avi17:064103,
    author = {Avila, Gustavo and Carrington, Tucker, Jr.},
    title = {Reducing the cost of using collocation to compute vibrational energy levels: Results for CH2NH},
    journal = jcp,
    volume = {147},
    number = {6},
    pages = {064103},
    year = {2017},
    month = {08}}

@article{wod19:154108,
  title={A pruned collocation-based multiconfiguration time-dependent Hartree approach using a Smolyak grid for solving the Schr{\"o}dinger equation with a general potential energy surface},
  author={Wodraszka, Robert and Carrington, Tucker},
  journal=jcp,
  volume={150},
  number={15},
  year={2019},
  pages={154108},
  publisher={AIP Publishing}
}

@article{wod21:114107,
    author = {Wodraszka, Robert and Carrington, Tucker, Jr.},
    title = {A rectangular collocation multi-configuration time-dependent Hartree (MCTDH) approach with time-independent points for calculations on general potential energy surfaces},
    journal = jcp,
    volume = {154},
    number = {11},
    pages = {114107},
    year = {2021},
    month = {03}}

@article{dra17:244108,
    author = {Dral, Pavlo O. and Owens, Alec and Yurchenko, Sergei N. and Thiel, Walter},
    title = {Structure-based sampling and self-correcting machine learning for accurate calculations of potential energy surfaces and vibrational levels},
    journal = jcp,
    volume = {146},
    number = {24},
    pages = {244108},
    year = {2017},
    month = {06}
}

@article{sch20:064105,
  title={A Gaussian process regression adaptive density guided approach for potential energy surface construction},
  author={Schmitz, Gunnar and Klinting, Emil Lund and Christiansen, Ove},
  journal=jcp,
  volume={153},
  number={6},
  pages={064105},
  year={2020},
  publisher={AIP Publishing}
}

@article{art20:194105,
  title={Adaptive density-guided approach to double incremental potential energy surface construction},
  author={Artiukhin, Denis G and Klinting, Emil Lund and K{\"o}nig, Carolin and Christiansen, Ove},
  journal=jcp,
  volume={152},
  number={19},
  year={2020},
  publisher={AIP Publishing}
}

@article{qu18:151,
  title={Permutationally invariant potential energy surfaces},
  author={Qu, Chen and Yu, Qi and Bowman, Joel M},
  journal=arpc,
  volume={69},
  number={1},
  pages={151--175},
  year={2018},
  publisher={Annual Reviews}
}

@article{hua25:035004,
  title={Active delta-learning for fast construction of interatomic potentials and stable molecular dynamics simulations},
  author={Huang, Yaohuang and Hou, Yi-Fan and Dral, Pavlo O},
  journal={Mach. Learn.: Sci. Technol.},
  volume={6},
  number={3},
  pages={035004},
  year={2025},
  publisher={IOP Publishing}
}

@article{sch23:144118,
  title={Positioning of grid points for spanning potential energy surfaces—How much effort is really needed?},
  author={Schneider, Moritz and Born, Daniel and K{\"a}stner, Johannes and Rauhut, Guntram},
  journal=jcp,
  volume={158},
  number={14},
  year={2023},
  pages={144118},
  publisher={AIP Publishing}
}

@article{pra13:69225,
  title={Ab initio potential energy and dipole moment surfaces for CS2: Determination of molecular vibrational energies},
  author={Pradhan, Ekadashi and Carreon-Macedo, Jose-Luis and Cuervo, Javier E and Schröder, Markus and Brown, Alex},
  journal=jpca,
  volume={117},
  number={32},
  pages={6925--6931},
  year={2013},
  publisher={ACS Publications}
}

@article{bro17:1730001,
  title={Fitting potential energy surfaces to sum-of-products form with neural networks using exponential neurons},
  author={Brown, Alex and Pradhan, E},
  journal=jtcc,
  volume={16},
  number={05},
  pages={1730001},
  year={2017},
  publisher={World Scientific}
}

@article{beh11:074106,
  title={Atom-centered symmetry functions for constructing high-dimensional neural network potentials},
  author={Behler, J{\"o}rg},
  journal=jcp,
  volume={134},
  number={7},
  year={2011},
  pages={074106},
  publisher={AIP Publishing}
}

@article{pan20:234110,
  title={Low-rank sum-of-products finite-basis-representation (SOP-FBR) of potential energy surfaces},
  author={Panad{\'e}s-Barrueta, Ram{\'o}n L and Pel{\'a}ez, Daniel},
  journal=jcp,
  volume={153},
  number={23},
  year={2020},
  pages={234110},
  publisher={AIP Publishing}
}

@article{nad23:114109,
  title={Analytical high-dimensional operators in canonical polyadic finite basis representation (CP-FBR)},
  author={Nadoveza, Nata{\v{s}}a and Panad{\'e}s-Barrueta, Ram{\'o}n L and Shi, Lei and Gatti, Fabien and Pel{\'a}ez, Daniel},
  journal=jcp,
  volume={158},
  number={11},
  year={2023},
  pages={114109},
  publisher={AIP Publishing}
}

\end{document}